\documentclass[epj]{webofc}
\usepackage[varg]{txfonts}   % Web of Conferences font
\usepackage{graphicx,color} % for figures
\usepackage{bm}       % bold math 
\usepackage{amsmath}  
\usepackage{amssymb}

\newcommand {\nc} {\newcommand}
\nc {\IR} [1]{\textcolor{red}{#1}}
\nc {\IB} [1]{\textcolor{blue}{#1}}
\nc {\IP} [1]{\textcolor{magenta}{#1}}
\woctitle{21st International Conference on Few-Body Problems in Physics}
\begin{document}
\title{Separable Forces for $(d,p)$ Reactions in Momentum Space}
\author{L.~Hlophe\inst{1} \and
        V.~Eremenko\inst{1,5}\fnsep \and
        Ch.~Elster\inst{1}\fnsep\thanks{\email{elster@ohio.edu}} \and
        F.~M.~Nunes\inst{2}\fnsep \and
        I.~J.~Thompson\inst{3}\fnsep \and
        G.~Arbanas\inst{4}\fnsep \and
        J.~E.~Escher\inst{3}\\[1mm]
        \textbf{TORUS} Collaboration${}^\dagger$
        (\url{http://reactiontheory.org})
        % etc.
}

\institute{Institute of Nuclear \& Particle Physics
  \textit{and} Dept. of Physics \& Astronomy,
  Ohio University, Athens, OH, 45710, USA
\and
  NSCL, Michigan State University, East Lansing, MI 48824, USA
\and
  Lawrence Livermore National Laboratory L-414, Livermore, CA 94551, USA
\and
  Nuclear Science and Technology Division, Oak Ridge National Laboratory,  Oak Ridge, TN
37831, USA
\and
  Institute of Nuclear Physics,
  Moscow State University, Moscow 119991, Russia
%  ${}^\dagger$\textit{Supported by U.S. Department of Energy,
%  Office of Science of Nuclear Physics}
}

\abstract{
Treating $(d,p)$ reactions in a Faddeev-AGS framework requires the interactions in the
sub-systems as input. We derived separable representations for the neutron- and
proton-nucleus interactions from phenomenological global optical potentials. In order to
take into account excitations of the nucleus, excitations need to be included explicity,
leading to a coupled-channel separable representation of the optical potential. 
}
\maketitle
%

%\vspace{-4mm}
%\section{Introduction}\label{intro}

 The $(d,p)$ scattering problem can be viewed as a three-body problem and thus be described
exactly by the Faddeev equations, which are more readily solved in momentum space. However, when considering
$(d,p)$ reactions involving heavier nuclei, currently employed screening techniques for solving Faddeev
equations with charged particles break down. At present, methods are developed to solve the Faddeev equations
in the Coulomb basis, however, those rely on the short range forces being separable. For 
the $np$ subsystem separable potentials are readily available in literature. However, 
this is not true for nucleon-nucleus interactions which are described by Woods-Saxon type
optical potentials. 

The method of deriving a separable representation of any arbitrary real potential proposed by Ernst,
Shakin, and Thaler (EST)~\cite{Ernst:1974zza} is well suited. EST separable 
potentials have the property that at specific chosen energies the wavefunctions 
corresponding to the original potential and its separable representation are identical. In order to
apply the EST method to optical potentials, it had to be generalized to non-Hermitian
potentials. Based on this generalized EST scheme neutron-nucleus optical
potentials for $^{48}$Ca, $^{132}$Sn, and $^{208}$Pb~\cite{lhlophe} were derived. Here a rank-5 
separable interactions were sufficient to provide a good description of the neutron-nucleus
scattering observables. 

\begin{figure}[ht]
\begin{minipage}[b]{0.49\linewidth}
%\centering
\includegraphics[width=0.95\linewidth]{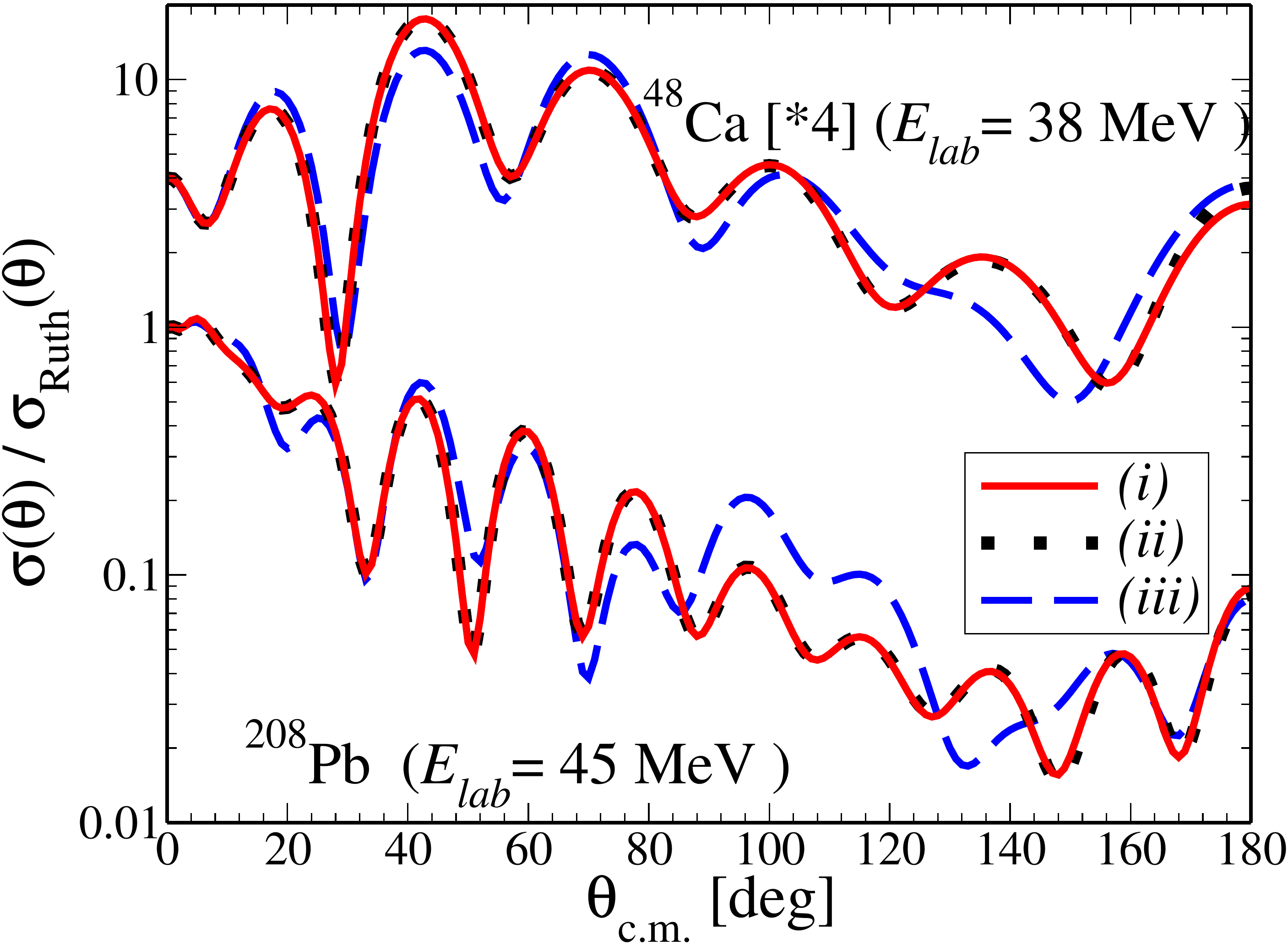}
\caption{\footnotesize{(color online) Unpolarized differential cross section for elastic scattering of 
protons from $^{48}$Ca (upper) and $^{208}$Pb (lower) divided by the Rutherford cross
section as function of the 
c.m. angle.  The solid lines ($i$) depict the cross
section calculated in momentum space based on the rank-5 separable representation of the
CH89~\cite{Varner:1991zz} optical potential,while the dotted lines
($ii$) represent the corresponding coordinate space calculations.
The dash-dotted lines
($iii$) show calculations in which the short-ranged Coulomb potential is omitted.\vspace{5mm}} 
}
\label{fig2}
\end{minipage}\hfill
\begin{minipage}[b]{0.47\linewidth}
%\centering
\includegraphics[width=0.98\linewidth]{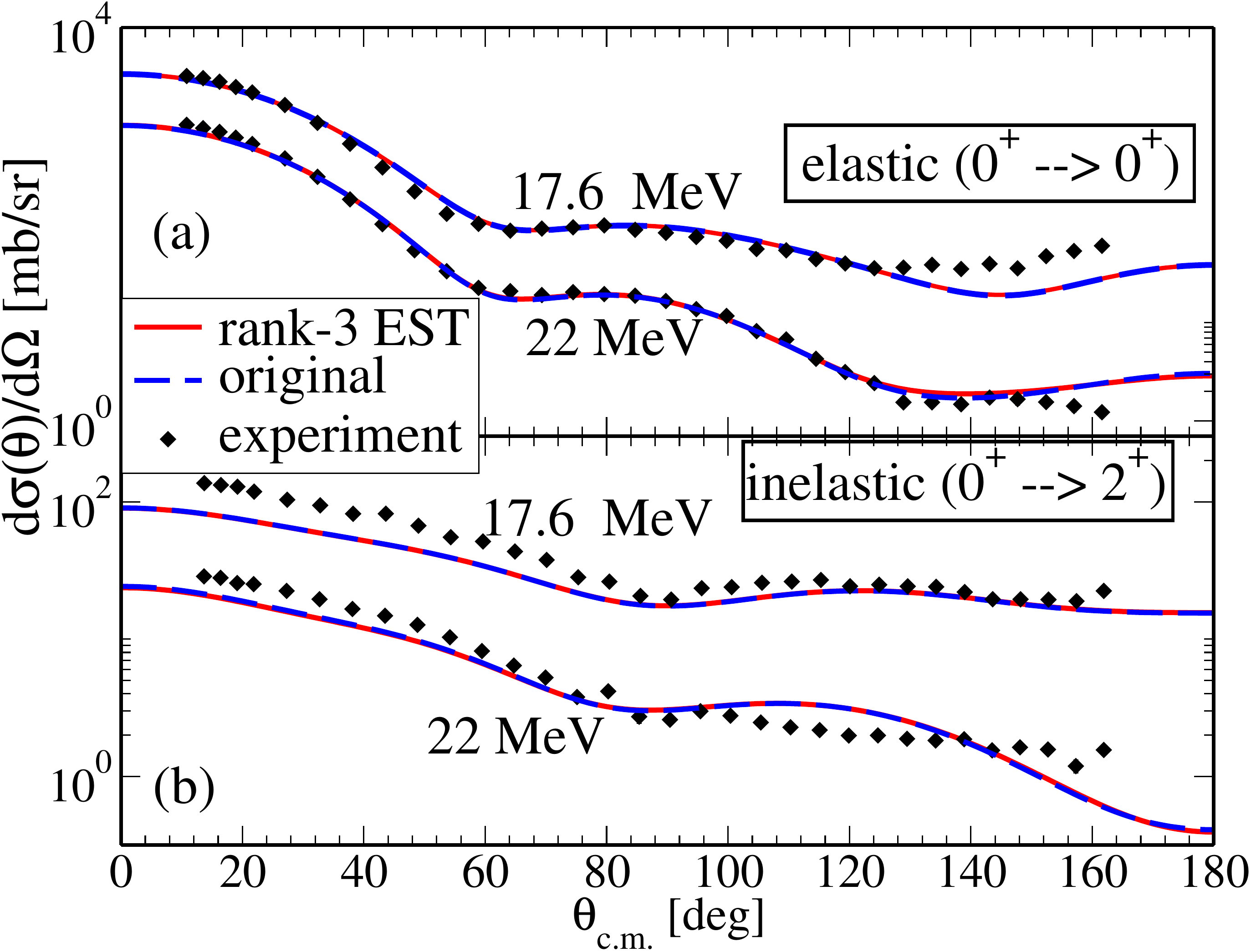}
\caption{\footnotesize{(color online) Unpolarized differential cross section for scattering of neutrons off $^{12}$C. 
Panel (a) shows the elastic scattering cross section while the inelastic $0^+\rightarrow 2^+$ 
cross section is shown in panel (b). The (blue) dashes show results obtained using the Olsson optical
potential~\cite{olsson89}. The (red) solid line indicates results obtained using a separable representation
of the Olsson optical potential with EST points at 5, 16.5, and 45 MeV. Experimental data from 
Ref.~\cite{olsson89} is depicted by black diamonds. The cross sections at 17.6 MeV are scaled  up by a factor
of four.}
}
\vspace{2mm}
\label{fig3}
\end{minipage}\hfill
%\begin{minipage}[b]{.3\linewidth}
%\includegraphics[width=6 cm]{figs/fig2.eps}

%\end{minipage}%\hfill

%\begin{minipage}[b]{0.3\linewidth}
%\includegraphics[width=6 cm]{figs/fig2.eps}

%\end{minipage}\hspace{1mm}%\hfill
\end{figure}

For the proton-nucleus optical potential one has to consider point Coulomb interaction, which is seen
at large distances, and a short range Coulomb potential describing the charged nuclear sphere. In order to extend the EST scheme to include the point Coulomb interaction,
one has to recast it in a Coulomb basis
instead of a plane wave basis. This extended EST scheme was then used to 
construct a separable representation of the proton-nucleus optical potential 
for $^{48}$Ca and $^{208}$Pb~\cite{lhlophec}. The cross-sections corresponding to the  CH89 proton-nucleus
phenomenological potential~\cite{Varner:1991zz} and its rank-5 separable representation are shown in Fig.~\ref{fig2}.
The good agreement between the two results illustrates the success of the EST scheme in
reproducing on-shell properties of the proton-nucleus optical potentials. 

In order to develop a potential for a deformed nucleus, non-spherical contributions (excitations) need
to be added.
A separable rank-3 representation of the $n+^{12}$C Olsson coupled-channels optical potential~\cite{olsson89}
including the 2$^+$ and 4$^+$ was constructed. The EST support points in the 0 to 50 MeV energy regime 
were determined to be 5, 16.5, and 45 MeV. In Fig.~\ref{fig3} elastic and inelastic scattering 
cross sections computed using the original Olsson potential (blue dashes) and its separable representation 
(red solid line) are shown.
Experimental data from Ref.~\cite{olsson89} is also presented. We observe that the quality of the 
coupled-channels separable representation matches that of the single channel case~\cite{lhlophe,lhlophec}.

%
% Non-BibTeX users please use
%

\begin{acknowledgement}
\vspace{-3mm}
This work was performed in part under the auspices of the
US Department of Energy, Office of Science of Nuclear Physics,
under the topical collaborations in nuclear theory program
No.~DE-SC0004087 (TORUS Collaboration), and under Contract DE-FG02-93ER40756 with Ohio
University.
\end{acknowledgement}

\end{document}